\documentclass[%
twoside,
reprint,
showpacs,
amsmath,amssymb,
aps,
pra,
]{revtex4-1}
\usepackage{graphicx}% Include figure files
\usepackage{dcolumn}% Align table columns on decimal point
\usepackage{bm}% bold math
\usepackage{hyperref}% add hypertext capabilities
\usepackage[mathlines]{lineno}% Enable numbering of text and display math
\usepackage[usenames]{color}
\usepackage{dsfont}
\usepackage{amssymb}
\usepackage[T1]{fontenc}
\usepackage{enumerate}
\usepackage{fancyhdr}
\usepackage{eqnarray,amsmath}\usepackage[usenames]{color}
\usepackage{relsize}
\usepackage{epstopdf}
\usepackage{subfigure}
\usepackage{float}
\usepackage{verbatim}
\usepackage{blkarray}
\usepackage{xcolor}

\begin{document}

\title{Exploring six modes of an optical parametric oscillator}

\author{Luis F. Mu\~n{}oz-Mart\'i{}nez\textsuperscript{1}}
\author{Felippe Alexandre Silva Barbosa\textsuperscript{2}}
\author{Ant\^onio Sales Coelho\textsuperscript{3,4}}
\author{Luis Ortiz-Guti\'e{}rrez\textsuperscript{5}}
\author{Marcelo Martinelli\textsuperscript{6,*}}
\author{Paulo Nussenzveig\textsuperscript{6}}
\author{Alessandro S. Villar\textsuperscript{7}}

\affiliation{
\textsuperscript{1} Departamento de F\'\i{}sica, Universidade Federal de Pernambuco, 50670-901 Recife, PE, Brazil  \\
\textsuperscript{2} Instituto de F\'\i{}sica Gleb Wataghin, Universidade Estadual de Campinas, 13083-859 Campinas, SP, Brasil\\
\textsuperscript{3} Departamento de Engenharia Mec\^anica, Universidade Federal do Piau\'\i{}, 64049-550 Teresina, PI, Brazil. \\
\textsuperscript{4}Departamento de F\'\i{}sica, Universidade Federal de Pernambuco, 50670-901 Recife, PE, Brazil  \\
\textsuperscript{5} Instituto de F\'\i{}sica de S\~a{}o Carlos, Universidade de S\~a{}o Paulo, P. O. Box 369, 13560-970 S\~a{}o Carlos, SP, Brazil\\
\textsuperscript{6} Instituto de F\'\i{}sica, Universidade de S\~a{}o Paulo, P.O.Box 66318, 05315-970 S\~a{}o Paulo, Brazil  \\
\textsuperscript{7} American Physical Society, 1 Research Road, Ridge, New York 11961, USA
}

\email{mmartine@if.usp.br}

\begin{abstract}
We measure the complete quantum state for six modes of the electromagnetic field produced by an optical parametric oscillator. The investigation involves the sidebands of the intense pump, signal, and idler fields generated by stimulated parametric downconversion inside a triply resonant optical resonator. We develop a theoretical model to successfully interpret the experimental results. The model takes into account the coupling of the field modes to the phonon bath of the nonlinear crystal, clearly showing the roles of different physical effects in shaping the structure of the quantum correlations between the six optical modes.
\end{abstract}

\maketitle

\section{Introduction}

The optical parametric oscillator has been used since the early days of quantum optics
to generate all sort of quantum states of light. 
The long list includes squeezed states \cite{kimblesqz}, intense twin beams \cite{twinfabre}, EPR entangled states \cite{pengentangled}, squeezed pump field \cite{pumpsqz}, entangled beams \cite{villarentangled}, three mode quantum correlations \cite{katiopexp} and three mode multicolor entanglement \cite{tripartite}.
The field modes produced by the optical parametric oscillator contain intricate quantum properties that are not yet completely understood, both in theory and in experiment.

The applications of these nonclassical states in the continuous variable domain goes from the use of squeezing for ultra-sensitive measurements \cite{caves, ligo} to the demands for entanglement in quantum information processing \cite{teleportkimble}, with convergence of experiments for discrete and continuous variables of the electromagnetic field \cite{teleportfurusawa}. Moreover, multimode entangled states in the continuous variable domain are interesting candidates for quantum information processing \cite{multimode}, leading to the search of sources  involving modes defined either in time \cite{multimodefurusawa}, frequency \cite{multimodefabre,multimodepfister} or momentum \cite{Wang2017}.

The fundamental process for the generation of these nonclassical states of light is the reversible exchange of energy among the pump field and the two downconverted modes.
With the aid of optical cavities, this effect is enhanced, and the output states can be calculated with the help of the input-output formalism for optical cavities and the master equation of the interacting Hamiltonian for the three modes of the field \cite{reiddrummond}.

Nevertheless, a detailed investigation over the detection process leads to a more complete description of the quantum state represented in the basis of field quadratures \cite{hexaopo}. In fact, optical detection is generally based in interferometric techniques, either by optical homodyning or by resonator self-homodyning \cite{prlsideband}. On the other hand, the measured quantum noise of light is analyzed in the frequency domain with the help of an electronic local oscillator to filter the contribution at a given frequency, associated to the sidebands of the optical field. Therefore, with careful data treatment, it can be shown that although the three mode description remains a valid approach, a more complete one can be obtained for the six detected modes of the field.

Our interest here is to present an explicit evaluation of the quantum state for the six sideband modes of the OPO that are measured by homodyne techniques and to access modal correlations that would not be available in the simplified three-mode picture of single-beam quantum fluctuations (pump, signal and idler). By explicitly using frequency modes of the field in the Hamiltonian,  we are able to deal with open cavities, looking for a more faithful description of optical setups usually involved in the nonclassical state generation. This six  mode description allows the complete analysis of entanglement in the OPO, demonstrating a deep hexapartite entangled structure for this system.
Moreover, the detailed sideband description puts in evidence the role of each field in the evolution of the system, something that  remained implicit in the usual treatment \cite{reiddrummond}. This allows the complete analysis of the hexapartite entanglement, to be treated in detail in another publication \cite{hexapartite}.

We begin by presenting the Hamiltonian for the sideband coupling in the nonlinear medium (Sec. \ref{hamilt}), and the evolution of the field operators under propagation on this medium (Sec. \ref{motion}). It is followed by the detailed model for the open cavity that is used to evaluate the operators of the output field (Sec. \ref{cavity}).
With the relation between the output and the input modes, momenta of any order can be evaluated. In the present scenario, we will limit the study to the second order momenta, and the reconstruction of the covariance matrix (Sec. \ref{covariance}).
Nevertheless, the description wouldn't be complete without the coupling of phonons to the sideband modes, included in the Hamiltonian of the system (Sec. \ref{phonon}).
The obtained results are used to describe the latest experimental results form our setup at different pump powers (Sec. \ref{experiment}), with pump powers up to 75\% above the oscillation threshold.
The complete description of the OPO in terms of the measured sidebands opens the possibility to analyze the multipartite entanglement present in this system in a wide range of operational conditions (Sec. \ref{conclusion}).

\section{\label{hamilt} Interaction Hamiltonian in the sidebands}

Each annihilation operator of the field $\hat{a}^{(n)}(t)$ is associated  to the electric field operator of a propagating wave and, in the limit of a cavity of infinite size, can be described by the contribution of operators at each frequency mode as \cite{quantumoptics}
\begin{eqnarray}\label{a2a}
\hat{a}^{(n)}(t)=e^{-i\omega_{n}t}\int_{-\omega_{n}}^{\infty} d\Omega e^{-i\Omega t} \hat{a}_{\omega_{n}+\Omega}^{(n)},
\end{eqnarray}
where $\hat{a}_{\omega_{n}+\Omega}$ 
is the photon annihilation operator in the mode of frequency $\omega=\omega_{n}+\Omega$, and we explicitly identify the carrier frequency of each field  ($\omega_{n}$)  and the frequency shift of each sideband  relative to this carrier $\Omega$. The mode $(n)$ specifies different directions of propagation, polarizations or carrier frequencies.

A usual treatment in optical systems considers as \textit{carrier} the mode with a significant population of photons, that is much larger than the average number of photons on all other modes. Therefore, in a linearized description of the fields by their mean value and a fluctuation, where each mode is described as $\hat{a}_{\omega_{n}+\Omega}^{(n)}=\langle \hat{a}_{\omega_{n}+\Omega}^{(n)} \rangle+\delta \hat{a}_{\omega_{n}+\Omega}^{(n)}$, we consider that  $|\alpha_{\omega_{n}}|^2\equiv \langle \hat{a}_{\omega_{n}}^{(n)\dag}\hat{a}_{\omega_{n}}^{(n)} \rangle \gg \langle \hat{a}_{\omega_{n}+\Omega}^{(n)\dag}\hat{a}_{\omega_{n}+\Omega}^{(n)} \rangle$ for $|\Omega|>\epsilon$, where $\alpha_{\omega_{n}}$ is the mean field of the carrier mode $(n)$ and $\epsilon$ is the carrier linewidth.

We can describe the interaction among the fields in a medium with a   second-order susceptibility $\chi$ with the help of an effective Hamiltonian 
\begin{eqnarray}\label{a1}
\hat{H}_{\chi}=i\hslash \dfrac{\chi}{\tau}\left[ \hat{a}^{(0)}(t)\hat{a}^{(1)\dagger}(t)
\hat{a}^{(2)\dagger}(t)-\text{h.c.}\right].
\end{eqnarray}
where $\tau$ is the time of flight through the medium and  field indexes 0, 1 and 2 stands for pump, signal and idler modes, respectively.

Using linearization,  we can rewrite this interaction Hamiltonian separating the contribution of each carrier and each sideband. In the triple product, only the terms satisfying energy conservation condition will prevail under propagation. This will include the relation for the carriers ($\omega_{0}=\omega_{1}+\omega_{2}$), as well as their sidebands.

This procedure will help to discriminate different contributions to the resulting Hamiltonian, coming from each mode involved. We will have the triple product of the carriers, associated to the mean value of the intense fields, as a constant value than can be disregarded for the evolution of operators. Next, there will be a combination of bilinear Hamiltonians for the specific sidebands shifted by $\pm \Omega$ from the central carriers
\begin{multline}
\hat{H}_{\chi}(\Omega)=-i\hslash \dfrac{\chi}{\tau}\Big[  \alpha_{\omega_{0}}^{*}\Big( \hat{a}^{(1)}_{\omega_{1}+\Omega} \hat{a}^{(2)}_{\omega_{2}-\Omega}+\hat{a}^{(1)}_{\omega_{1}-\Omega}\hat{a}^{(2)}_{\omega_{2}+\Omega}\Big)+
\\
 \alpha_{\omega_{1}}\left( \hat{a}^{(0)\dagger}_{\omega_{0}+\Omega}\hat{a}^{(2)}_{\omega_{2}+\Omega} +\hat{a}^{(0)\dagger}_{\omega_{0}-\Omega}\hat{a}^{(2)}_{\omega_{2}-\Omega}\right)+
 \\
 \alpha_{\omega_{2}}\left( \hat{a}^{(0)\dagger}_{\omega_{0}+\Omega} \hat{a}^{(1)}_{\omega_{1}+\Omega}+\hat{a}^{(0)\dagger}_{\omega_{0}-\Omega}\hat{a}^{(1)}_{\omega_{1}-\Omega}\right) -\text{h.c.} \Big].
\label{a6}
\end{multline}
defined for $ \Omega > \epsilon$ for convenience.
Linear terms on the fluctuations will not satisfy  energy conservation, and contribution of tri-linear or cubic terms will be negligibly small in comparison with the bi-linear terms involving the intense mean fields of the carriers, and will be disregarded in the present treatment.
Hamiltonian in Eq. (\ref{a1}) may be described by the sum over the contribution of each Hamiltonian from Eq.(\ref{a6}) for different frequencies $\Omega$,
$\hat{H}_{\chi}=\int_{\epsilon}^{\infty} \hat{H}_{\chi}(\Omega) \, d\Omega$.
Therefore, under the validity of linearization, each set of sideband pairs defined by $\Omega> \epsilon$ is decoupled from other sets defined by $\Omega'\neq \Omega$.

On the other hand, upper and lower sidebands are coupled in pairs in Eq. (\ref{a6}). The field operators of these sidebands are pairwise measured by the treatment of detected photocurrents in the frequency domain \cite{prlsideband,hexaopo}. The treatment for the evolution of these operators can be simplified if we change to the measurement basis involving symmetric ($\mathcal{S}$) and antisymmetric ($\mathcal{A}$) combinations of upper and lower  sidebands operators \cite{hexaopo}
\begin{eqnarray}\label{aa6a}
\hat{a}^{(n)}_{s(a)}=\dfrac{1}{\sqrt{2}}\Big[ \hat{a}^{(n)}_{\omega_{n}+\Omega}\pm \hat{a}^{(n)}_{\omega_{n}-\Omega}\Big].
\end{eqnarray}
On this basis, the Hamiltonian given in Eq. (\ref{a6}) is rewritten as
\begin{eqnarray}\label{aa8}
\hat{H}_{\chi}(\Omega)=\hat{H}_{\chi s}+\hat{H}_{\chi a},
\end{eqnarray}
where
\begin{multline}\label{aa8b}
\hat{H}_{\chi s(a)}=-i\hslash\dfrac{\chi}{\tau}\Big[ \pm \alpha_{\omega_{0}}^{*}\hat{a}^{(1)}_{s(a)} \hat{a}^{(2)}_{s(a)}+\\
\alpha_{\omega_{1}}\hat{a}^{(0)\dagger}_{s(a)}\hat{a}^{(2)}_{s(a)}+
\alpha_{\omega_{2}}\hat{a}^{(0)\dagger}_{s(a)} \hat{a}^{(1)}_{s(a)}- \text{h.c.}\Big],
\end{multline}
where the $+$ ($-$) signal is used for the symmetric (antisymmetric) combination of sidebands along this article.
This Hamiltonian describes a process leading to two-mode squeezing involving downconverted modes $\hat{a}^{(1)}_{s(a)}$ and $\hat{a}^{(2)}_{s(a)}$ mediated by the intense pump field, and two beam splitter processes exchanging photons between the pump and each downconverted mode, mediated by the intense complementary downconverted field. These three process lead to a rich entanglement dynamics, that was understood as a source of tripartite entangled fields in the symmetric mode description \cite{tripartiteprl}.
Beyond this three mode description, a rich mesh of entanglement dynamics involving six modes is generated by Eq. (\ref{a6}), combining creation and annihilation of pairs of photons in downconverted sidebands and photon exchange between pump and downconverted sidebands, leading to hexapartite entanglement among the involved modes \cite{hexapartite}.

On the other hand, Eq. (\ref{aa8b}) shows that the subspaces of symmetric and antisymmetric combinations of sidebands are not coupled by the nonlinear medium. Nevertheless, these correlations were already observed in experiments \cite{hexaopo}, and its origin is found somewhere else in the OPO, as we will see in Sec. (\ref{cavity}).

\section{Equations of Motion and solution by matrix method\label{motion}}

After passing through the nonlinear medium, the modes in subspaces of $\mathcal{S}/\mathcal{A}$ combinations of sidebands will interact according to the Hamiltonian given by Eq.(\ref{aa8}). Therefore, the equations describing the evolution of the operators during their
propagation through the medium are given by
\begin{eqnarray}
\dfrac{d \hat{a}^{(0)}_{s(a)}}{d \xi}&=&-\chi\Big[\alpha_{\omega_{1}} \hat{a}^{(2)}_{s(a)}+ \alpha_{\omega_{2}} \hat{a}^{(1)}_{s(a)}\Big]\label{b10}\\
\dfrac{d \hat{a}^{(1)}_{s(a)}}{d \xi}&=&\chi\Big[
\pm \alpha_{\omega_{0}}\hat{a}^{(2)\dagger}_{s(a)} + \alpha_{\omega_{2}}^{*}\hat{a}^{(0)}_{s(a)}\Big]\label{b11}\\
\dfrac{d \hat{a}^{(2)}_{s(a)}}{d \xi}&=&\chi\Big[\pm\alpha_{\omega_{0}} \hat{a}^{(1)\dagger}_{s(a)}+ \alpha_{\omega_{1}}^{*} \hat{a}^{(0)}_{s(a)}\Big],\label{b12}
\end{eqnarray}
where 
 $\xi$ the normalized time evolution given by $\xi=t/\tau$.

Defining $\vec{\mathbf{A}}_{s(a)}=(\hat{a}^{(0)}_{s(a)} ~\hat{a}^{(0)\dagger}_{s(a)} ~\hat{a}^{(1)}_{s(a)} ~\hat{a}^{(1)\dagger}_{s(a)} ~\hat{a}^{(2)}_{s(a)} ~\hat{a}^{(2)\dagger}_{s(a)})^{T}$, the set of differential equations given by Eqs. (\ref{b10}-\ref{b12}) and their Hermitian adjoints can be written as 
\begin{eqnarray}\label{m2}
\dfrac{d \vec{\mathbf{A}}_{s(a)}}{d \xi}=\mathbf{M}_{\chi s(a)} \vec{\mathbf{A}}_{s(a)},
\end{eqnarray}
where
\begin{eqnarray*}\label{matrix1}
\mathbf{M}_{\chi s(a)}=\chi\left(
\begin{matrix}
0 & 0 & -\alpha_{\omega_{2}} & 0 & -\alpha_{\omega_{1}} & 0 \\
0 & 0 & 0 & -\alpha_{\omega_{2}}^{*} & 0 & -\alpha_{\omega_{1}}^{*} \\
\alpha_{\omega_{2}}^{*} & 0 & 0 & 0 & 0 & \pm \alpha_{\omega_{0}} \\
0 & \alpha_{\omega_{2}} & 0 & 0 & \pm \alpha_{\omega_{0}}^{*} & 0 \\
\alpha_{\omega_{1}}^{*} & 0 & 0 & \pm \alpha_{\omega_{0}} & 0 & 0 \\
0 & \alpha_{\omega_{1}} & \pm \alpha_{\omega_{0}}^{*} & 0 & 0 & 0
\end{matrix}\label{coupling matrix}
\right).
\end{eqnarray*}
From Eq.  (\ref{m2}) the field leaving the crystal can be written as
\begin{eqnarray}\label{m4}
\vec{\mathbf{A}}_{s(a)}\Big| _{\xi=1}=\mathbf{G}_{s(a)}(\chi) \vec{\mathbf{A}}_{s(a)}\Big|_{\xi=0},
\end{eqnarray}
where
\begin{eqnarray}
\mathbf{G}_{s(a)}(\chi)=\exp\left(\int_{0}^{1}d\xi~\mathbf{M}_{\chi s(a)} \right).\label{gain}
\end{eqnarray}
The matrix $\mathbf{G}_{s(a)}(\chi)$ is defined as the \textit{gain matrix of the medium}, and
allows the evaluation of  all $\hat{a}^{(n)}_{\omega_{n}\pm \Omega}$ and their Hermitian adjoints after passing through the crystal.

In the calculation of the evolution of the terms inside the cavity, it will be useful to play with all creation and annihilation operators of the involved sidebands in a vector form $\vec{\mathbf{A}}=(\hat{a}^{(0)}_{\omega_{0}+\Omega} ~\hat{a}^{(0)\dagger}_{\omega_{0}+\Omega}~\cdots ~\hat{a}^{(0)}_{\omega_{0}-\Omega} ~\hat{a}^{(0)\dagger}_{\omega_{0}-\Omega}~\cdots )^{T}$,
 related to vectors $\vec{\mathbf{A}}_{s(a)}$ as% follows
\begin{eqnarray}\label{m6}
 \vec{\mathbf{A}}=\mathbf{\Lambda} \left(\vec{\mathbf{A}}_{s}, \vec{\mathbf{A}}_{a}  \right)^{T},
\end{eqnarray}
where the transformation matrix is of the form
\begin{eqnarray}\label{transform}
\mathbf{\Lambda}=\mathbf{\Lambda}^{-1}=\dfrac{1}{\sqrt{2}}
\left(
\begin{tabular}{cccccccccccc}
$\mathbf{1}_{6\times6}$ & $\mathbf{1}_{6\times6}$  \\
$\mathbf{1}_{6\times6}$ & -$\mathbf{1}_{6\times6}$
\end{tabular}
\right),
\end{eqnarray}
where $\mathbf{1}_{6\times6}$ are identity matrices of order 6.
Taking into account Eqs. (\ref{m4}) and (\ref{m6}), the transformation of the field operators that propagated through the medium is given by
\begin{eqnarray}\label{m8}
 \vec{\mathbf{A}}\Big|_{\xi=1}=\mathbf{G}(\chi) \vec{\mathbf{A}}\Big|_{\xi=0},
\end{eqnarray}
where
\begin{eqnarray}\label{m9}
 \mathbf{G}(\chi)=\mathbf{\Lambda}\left(\mathbf{G}_{s}(\chi)\oplus \mathbf{G}_{a}(\chi)\right)\mathbf{\Lambda}.
\end{eqnarray}
The symbol $\oplus$ represents a direct sum, resulting in a block diagonal matrix.

Thanks to the bilinear form of the Hamiltonian in Eq. (\ref{a6}), we have a linear evolution of the coupling of different fields through the medium, that will contribute to the equations describing their evolution inside a cavity.

\section{Physical effect of the optical cavity \label{cavity}}

\begin{figure}[t]
  \begin{center}
    \includegraphics[width=0.7\linewidth]{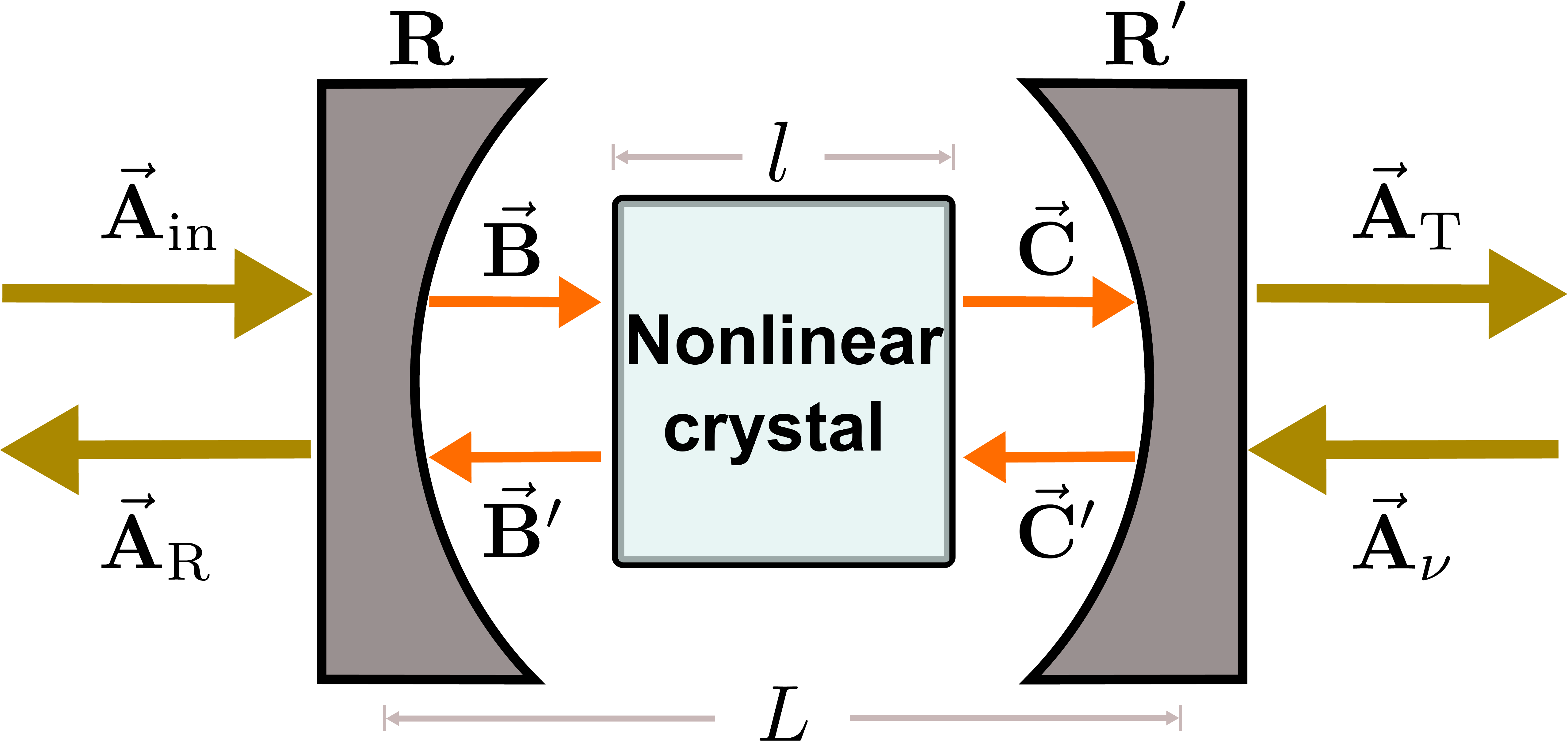}
    \caption{Basic configuration of OPO, consisting of a nonlinear medium of length $l$ inside a linear cavity of lenght $L$, made of one coupling mirror (left) and one end mirror (right) accounting for spurious losses. }
    \label{fig:OPO}
  \end{center}
\end{figure}

It must be kept in mind that our goal is to theoretically model the evolution of the sideband modes of an OPO, consisting of a nonlinear crystal located in a linear cavity that we assume to have arbitrary losses for the fields involved,
as described in Fig. \ref{fig:OPO}.
 The coupling mirror has  reflection and transmission coefficients, $r_{n}$ and $t_{n}$, for each carrier, and the end mirror, with reflection coefficient $r'_{n}$ and transmission coefficient $t'_{n}$, accounts for  spurious losses (that may include absorption in the crystal or scattering on the optical interfaces).
 These coefficients can be conveniently described by loss parameters $\gamma_{n}$ and
 $\gamma'_{n}$ as
\begin{eqnarray}
r_{n}= e^{-\gamma_{n}},
\quad\quad
t_{n}= (1-r'^{2}_{n})^{1/2},\nonumber\\
r'_{n}= e^{-\gamma'_{n}},
\quad\quad
t'_{n}= (1-r'^{2}_{n})^{1/2}\label{m19a}.
\end{eqnarray}
The total loss in a round trip can be directly evaluated from $\gamma^t_{n}=\gamma_{n}+\gamma'_{n}$.
%The formalism adopted here  remains valid even in the open cavity regime, enabling the treatment in the limit where the cavity is completely open for one of the modes, as in the case of a doubly resonant OPO \cite{opencavity}.

The equations relating each field operator inside and outside the cavity (Fig. \ref{fig:OPO}) are given by the beam splitter transformation
\begin{align}
\vec{\mathbf{A}}_{\textsc{R}}&= \textbf{R}\vec{\mathbf{A}}_{\text{in}}+\textbf{T}\vec{\mathbf{B}}', &
\vec{\mathbf{B}}&= \textbf{T}\vec{\mathbf{A}}_{\textrm{in}}- \textbf{R}\vec{\mathbf{B}}',&\label{m13}\\[2mm]
\vec{\mathbf{A}}_{\textrm{T}}&= \textbf{R}'\vec{\mathbf{A}}_{\nu}+\textbf{T}' \vec{\mathbf{C}},&
\vec{\mathbf{C}}'&= \textbf{T}'\vec{\mathbf{A}}_{\nu}- \textbf{R}'\vec{\mathbf{C}},&\label{m14}
\end{align}
with
\begin{align}
\label{rt}
\textbf{R}&=\text{diag}\big(r_{0}~r_{0}~r_{1}~r_{1}~r_{2}~r_{2}~ r_{0}~r_{0}\cdots\big),\notag\\[2mm]
\textbf{T}&=\text{diag}\big(t_{0}~t_{0}~t_{1}~t_{1}~t_{2}~t_{2}~ t_{0}~t_{0}\cdots\big),\notag\\[2mm]
\textbf{R}'&=\text{diag}\big(r'_{0}~r'_{0}~r'_{1}~r'_{1}~r'_{2}~r'_{2}~ r'_{0}~r'_{0}\cdots\big),  \notag\\[2mm]
\textbf{T}'&=\text{diag}\big(t'_{0}~t'_{0}~t'_{1}~t'_{1}~t'_{2}~t'_{2}~ t'_{0}~t'_{0}\cdots\big),
\end{align}
keeping the vector ordering for the field operators we used in the previous section.
The fields described by $\vec{\mathbf{A}}_{\text{in}}$ enters the cavity through the coupling mirror, while
$\vec{\mathbf{A}}_{\nu}$ models the fields associated to vacuum modes coupled through spurious losses.

Each field $\hat{a}^{(n)}_{\omega_{n}\pm\Omega}$ will be transformed by the gain inside the crystal as described by Eq. (\ref{m8}). Besides that, their phase will evolve during the propagation along the cavity.
Under perfect phase matching conditions \cite{Debuisschert93}, if the refractive index for the fields are close enough, we may consider that  the evolution of the phase commutes with the gain. Therefore the relation between the propagating fields on each side of the cavity will be given by
\begin{eqnarray}
\vec{\mathbf{C}}= e^{-i \boldsymbol{\varphi}}\mathbf{G}(\chi) \vec{\mathbf{B}}, \quad \quad
\vec{\mathbf{B}}'= e^{-i\boldsymbol{\varphi}}\mathbf{G}(\chi)  \vec{\mathbf{C}}'.\label{m19}
\end{eqnarray}
The phase vector
\begin{equation}
\boldsymbol{\varphi}=\boldsymbol{\varphi}(\Omega) \oplus \boldsymbol{\varphi}(-\Omega),\label{m20}
\end{equation}
with,
\begin{eqnarray*}
\boldsymbol{\varphi}(\Omega)=  \text{diag}\big(
\varphi_{\Omega}^{(0)},-\varphi_{\Omega}^{(0)}
\varphi_{\Omega}^{(1)}, -\varphi_{\Omega}^{(1)},
\varphi_{\Omega}^{(2)}-\varphi_{\Omega}^{(2)} \big),
\end{eqnarray*}
gives a different contribution for each sideband depending of the frequency shift $\Omega$ and on the carrier frequency $\omega_n$
\begin{eqnarray}
\boldsymbol{\varphi}_{\Omega}^{(n)}=
\dfrac{\omega_{n}+\Omega}{2\,\text{FSR}_{n}}.
\end{eqnarray}
where $\text{FSR}_{n}=c/2L_{\text{op}}^{(n)}$, is the free spectral range for the mode $n$, with
$L_{\text{op}}^{(n)}=L+l(\verb"n"_{n}-1)$ being the effective optical length between the cavity mirrors, depending on the crystal refractive index $\verb"n"_{n}$, and on  the speed of light $c$.
Evidently the effective phase contribution will depend on the detuning between the carrier and the nearest cavity mode $\omega^c_n$, an integer multiple of $2\pi \text{FSR}_{n}$, given by  $\Delta_n=\omega_n -\omega^c_n$.

An important point related to the evolution of the sidebands should be noticed. Each operator will undergo a different phase evolution, depending on their frequency. That will mix symmetric and antisymmetric modes, even for null carrier detuning, since upper and lower sidebands will, in this case, undergo opposite phase evolutions. This is the cause of the correlations between symmetric and antisymmetric modes observed in \cite{hexaopo}.

Combining beam splitter transformation, phase evolution and gain, expressed in Eqs.(\ref{m13}--\ref{m19}) we can derive a linear transformation for the reflected modes, coupled to the incident modes on the OPO, as
\begin{eqnarray}
\vec{\mathbf{A}}_{\textrm{R}}=  \textbf{R}_{\chi}\vec{\mathbf{A}}_{\textrm{in}}+\textbf{T}'_{\chi}
\vec{\mathbf{A}}_{\nu},\label{ar24}
\end{eqnarray}
where
\begin{eqnarray}
\textbf{R}_{\chi}&=&\textbf{R}-\textbf{T}e^{-i\boldsymbol{\varphi}}
\mathbf{G}(\chi)\textbf{R}'e^{-i\boldsymbol{\varphi}}\mathbf{G}(\chi)
\mathbf{D}(\chi)\textbf{T},\label{Eq:26}\\[2mm]
\textbf{T}'_{\chi}&=& \textbf{T}e^{-i\boldsymbol{\varphi}}\mathbf{G}(\chi) \left[\textbf{I} +\textbf{R}'
e^{-i\boldsymbol{\varphi}}\mathbf{G}(\chi)\mathbf{D}(\chi)
\textbf{R}e^{-i\boldsymbol{\varphi}}\mathbf{G}(\chi) \right] \textbf{T}',\nonumber\\\label{Eq:27}
\end{eqnarray}
and
\begin{eqnarray}
\mathbf{D}(\chi)=\Big( \textbf{1}-\textbf{R}e^{-i\boldsymbol{\varphi}}\mathbf{G}(\chi) \textbf{R}'e^{-i\boldsymbol{\varphi}}\mathbf{G}(\chi)\Big)^{-1}.\label{Eq:28}
\end{eqnarray}

We should notice that the conversion matrix given by Eq. (\ref{transform}), relating individual modes to symmetric/antisymmetric combination, commutes with the reflection and transmission matrices given by Eq. (\ref{rt}), but not with the phase evolution matrix.
It is consistent with the fact that the coupling of symmetric and antisymmetric modes comes from the opposite phase evolution for the sidebands.
Another interesting point of the formalism here adopted is that it allows the evaluation of the complete covariance matrix for the sideband modes, in an approach valid for lossy cavities beyond the narrowband regime employed in ref. \cite{collet}. In the extreme limit, it could be used to study the transformation of field in doubly resonant cavities, even for the mode undergoing a single pass through the nonlinear medium.

\section{Hexapartite quantum state\label{covariance}}

Consistent with the description used in ref.\cite{hexaopo},
we can evaluate the covariance matrix for the field quadratures $\hat{p}^{(n)}_{\omega}$ and $\hat{q}^{(n)}_{\omega}$ related to photon annihilation $\hat{a}^{(n)}_{\omega}$ operator as  $\hat{a}^{(n)}_{\omega}=(\hat{p}^{(n)}_{\omega}+i \hat{q}^{(n)}_{\omega})/2$ and
satisfying the commutation relation $[\hat{p}^{(n)}_{\omega},\hat{q}^{(n)}_{\omega '}]=2 i \delta(\omega-\omega ')$.
The relevant quadrature operators  can be ordered in a column vector $\vec{\mathbf{X}}=(\hat{p}^{(0)}_{\omega}~ \hat{q}^{(0)}_{\omega}~\cdots ~ \hat{p}^{(n)}_{\omega'}~\hat{q}^{(n)}_{\omega'} \cdots)^{T}$, that is directly related to the vector of field operators by $\vec{\mathbf{X}}=\textbf{N}\vec{\mathbf{A}}$.

Second order momenta of the field operators are all contained in the symmetrized covariance matrix, evaluated over the quantum state of the system as
\begin{eqnarray}
\mathbf{V}=\frac{1}{2}\left( \langle \vec{\mathbf{X}}\cdot \vec{\mathbf{X}}^{T} \rangle + \langle \vec{\mathbf{X}}\cdot \vec{\mathbf{X}}^{T} \rangle^{T}\right).
\label{V}
\end{eqnarray}
Diagonal elements of $\mathbf{V}$ represent variances of single-mode quadrature operators, denoted as, e.g., $\Delta^{2}\hat{p}^{(n)}_{\omega}\equiv \langle \hat{p}^{(n)}_{\omega} \hat{p}^{(n)}_{\omega}\rangle$. Off-diagonal elements are correlations between different quadratures operators, such as in, e.g., $C(\hat{p}^{(n)}_{\omega}\hat{p}^{(m)}_{\omega'})\equiv(\langle \hat{p}^{(n)}_{\omega}\hat{p}^{(m)}_{\omega'}\rangle + \langle \hat{p}^{(m)}_{\omega'}\hat{p}^{(n)}_{\omega} \rangle)/2$.

The basis transformation given by matrix $\textbf{N}$ applied to Eq. (\ref{ar24}) results in quadrature
operators  $\vec{\mathbf{X}}_{\textrm{R}}=  \widetilde{\textbf{R}}_{\chi}\vec{\mathbf{X}}_{\textrm{in}}
+\widetilde{\textbf{T}}'_{\chi}\vec{\mathbf{X}}_{\nu}$, where $\widetilde{\textbf{R}}_{\chi}=\textbf{N} \textbf{R}_{\chi} \textbf{N}^{-1}$, $\widetilde{\textbf{T}}'_{\chi}=\textbf{N} \textbf{T}'_{\chi} \textbf{N}^{-1}$. Thus, the evaluation of the covariance matrix for the output fields results in
\begin{equation}
\mathbf{V}_{\textrm{R}}=\widetilde{\textbf{R}}_{\chi}\mathbf{V}_{\textrm{in}}
\widetilde{\textbf{R}}_{\chi}^{T}+ \widetilde{\textbf{T}}'_{\chi}\mathbf{V}_{\nu}\widetilde{\textbf{T}}_{\chi}'^{T},
\label{Eq:30}
\end{equation}
where $\mathbf{V}_{\textrm{in}}$ is the input field covariance matrix and $\mathbf{V}_{\nu}$ is the covariance matrix of the field entering through the cavity loss channels.
For losses coupling the cavity to vacuum modes  we have $\mathbf{V}_{\nu}=\textbf{1}$.

The covariance matrix in the basis of $\mathcal{S}/\mathcal{A}$ combinations of sidebands will have the same form described in ref. \cite{hexaopo}
\begin{eqnarray}
\mathbb{\mathbf{V}}_{\textrm{R}(s/a)}=
\begin{pmatrix}
            \mathbf{V}_{s} & \mathbf{C}_{s/a} \\
            (\mathbf{C}_{s/a})^{T} & \mathbf{V}_{a}
\end{pmatrix}.
\end{eqnarray}
It is important to notice that the elements in the covariance matrices $\mathbf{V}_{s}$ and $\mathbf{V}_{a}$ are related by a $\pi/2$ rotation on the quadrature phase space, changing $\hat{p}_s\rightarrow \hat{q}_a$ and $\hat{q}_s\rightarrow -\hat{p}_a$ in covariance terms (e.g., $C(\hat{p}^{(n)}_{s}\hat{q}^{(m)}_{s})=-C(\hat{q}^{(n)}_{a}\hat{p}^{(m)}_{a})$, $\Delta^{2}\hat{p}^{(n)}_{s}=\Delta^{2}\hat{q}^{(n)}_{a}$,...). Therefore, the modeling described here is equivalent to the semiclassical approach often used in evaluation of the noise spectra with the help of Langevin equations \cite{collet, quantumoptics, phononpra, hexaopo}, and both methods can be used to obtain the same amount of information about the 2n modes of sidebands for n modes of carriers. 
However, it is important to clarify that the method developed here is explicit in presenting the physical origin of the correlations between symmetric and antisymmetric modes, something that was elusive in the semiclassical model. As demonstrated in Secs. (\ref{hamilt}) and (\ref{cavity}), these correlations are not generated only by the cavity, or by the squeezing generating term in Eq. (\ref{aa8b}), that is the only remaining term for operation below the oscillation threshold. It is their combination with the beam splitting term, associated to signal and idler mean fields, that will lead to these correlations.

Considering the particular case where the input is also a coherent state ($\mathbf{V}_{\textrm{in}}=\textbf{1}$), for exact resonance of the carriers ($\Delta_n=0$), we have
\begin{align}
\mathbf{V}_{s}=
\left(
  \begin{array}{cccccc}
    \rho^{(0)} & 0 & \mu^{(01)} & 0 & \mu^{(02)} & 0 \\
    0 & \beta^{(0)} & 0 & \nu^{(01)} & 0 & \nu^{(02)} \\
    \mu^{(01)} & 0 & \rho^{(1)} & 0 & \zeta^{(12)} & 0 \\
    0 & \nu^{(01)} & 0 & \beta^{(1)} & 0 & \epsilon^{(12)} \\
    \mu^{(02)} & 0 & \zeta^{(12)} & 0 & \rho^{(2)} & 0 \\
    0 & \nu^{(02)} & 0 & \epsilon^{(12)} & 0 & \beta^{(2)}
  \end{array}
  \right),\label{covarsphT1}
\end{align}
with 12 independent terms and
\begin{align}
\mathbf{C}_{s/a}=
\left(
  \begin{array}{cccccc}
    0 & 0 & 0 & -\kappa^{(01)} & 0 & -\kappa^{(02)} \\
    0 & 0 & \lambda^{(01)} & 0 & \lambda^{(02)} & 0 \\
    0 & \kappa^{(01)} & 0 & 0 & 0 & -\varrho^{(12)} \\
    -\lambda^{(01)} & 0 & 0 & 0 & \eta^{(12)} & 0 \\
    0 & \kappa^{(02)} & 0 & \varrho^{(12)} & 0 & 0 \\
    -\lambda^{(02)} & 0 & -\eta^{(12)} & 0 & 0 & 0
  \end{array}
  \right),\label{covarsphT2}
\end{align}
with 6 independent terms.

 Evaluation of the covariance matrix depends on the value of the mean fields, as can be seen in Eq.(\ref{coupling matrix}). If we go beyond the linearized model presented in ref. \cite{Debuisschert93}, the contributions to the gain matrix can be explicitly scaled to the oscillation threshold $|\alpha_{\omega_{0}}^{\text{in}}| ^{2}_{\text{th}}$ as
\begin{eqnarray*}
\chi^2\mid\alpha_{\omega_{0}}\mid^{2}&=&\frac{\left(1-e^{-2\gamma{}_{0}}\right)}{\left(1-e^{-\gamma^{t}{}_{0}}\right)^{2}}\, \chi^2\mid\alpha_{\omega_{0}}^{\text{{in}}}\mid_{\text{{th}}}^{2},\\
\chi^2 \mid\alpha_{\omega_{j}}\mid^{2}&=&\frac{e^{2\gamma^{t}{}_{0}}\left(1-e^{-2\gamma{}_{0}}\right)\left(\sqrt{\sigma}-1\right)}{\left(e^{\gamma^{t}{}_{0}}-1\right)\left(e^{\gamma^{t}{}_{j}}-1\right)}\, \chi^2\mid\alpha_{\omega_{0}}^{\text{{in}}}\mid_{\text{{th}}}^{2}
\label{fields12},\\
\end{eqnarray*}
with $j=1,2$, where the normalized pump power is given by
$\sigma=|\alpha_{\omega_{0}}^{\text{in}}| ^{2}/|\alpha_{\omega_{0}}^{\text{in}}| ^{2}_{\text{th}}$.
Moreover,
\begin{eqnarray*}
\chi^2 {\mid\alpha_{\omega_{0}}^{\text{{in}}}\mid_{\text{{th}}}^{2}}&=&\dfrac{\left(1-e^{-\gamma^{t}{}_{0}}\right)^{2}\left(e^{\gamma^{t}{}_{1}}-1\right)\left(e^{\gamma^{t}{}_{2}}-1\right)}{4\left(1-e^{-2\gamma{}_{0}}\right)},\label{fields12b}
\end{eqnarray*}
implies that all the mean values can be related only to the cavity coupling terms and the normalized pump power.

We have retained here the consideration that evolution of the mean field amplitude inside the crystal is negligible, as it was done in ref. \cite{Debuisschert93}. 
Further development can be done if we consider that these fields vary along the crystal. %cite{opencavity}.
Nevertheless, in the integration in Eq. (\ref{gain}), we see that their evolution
will not affect the linearity of the solution regarding the mode operators, and an effective contribution can be evaluated to obtain a precise description of the resulting covariances.

While this treatment could account for the OPO spectra above the threshold, is doesn't account for extra noise sources, as the phonon-photon coupling in the crystal \cite{phononpra}. Its effect can be included in the interaction Hamiltonian, as we will see next.
This extra phonon noise may also introduce correlations between $\hat{p}$ and $\hat{q}$ quadratures within $ \mathbf{V}_{s(a)}$ matrices, as well as correlations in $\mathbf{C}_{s/a}$ matrix, that can be also found in the case of non-zero cavity detunings.

\section{Physical effect of phonons in the nonlinear crystal in the quantum noise of light \label{phonon}}

In many experiments with above threshold OPO's, an extra phase noise appears on the optical fields which is caused by the scattering of light by thermal phonons within the crystal and which considerably modifies the quantum state of the system. A detailed semi-classical analysis of this effect was realized in ref. \cite{phononpra}. In this section we are going to establish a quantum model for this excess phase noise in order to have a consistent and complete quantum description of an OPO operating above threshold.

\subsection{Complete interaction Hamiltonian}

Photons that circulate inside the optical cavity of an OPO may eventually exert a small radiation pressure on the crystal,
leading to local density fluctuations associated with acoustic phonons. 
 On the other hand, fluctuations of the refractive index, of optical or mechanical origin, will result  in small phase fluctuations, leading to Stokes and Brillouin light scattering \cite{Boyd95} with frequency shifts in the scattered light.
 This process can also be seen as a random
 detuning of the optical cavity since it modifies its optical length  $L_{\text{op}}^{(n)}$.

 In the present case, we will be interested in the fraction of the scattering that is coupled to the cavity modes, with small shifts in the frequency (within the cavity bandwidth). The Hamiltonian which correctly models this type of photon-phonon interaction is known as optomechanical Hamiltonian \cite{optomecHamilt}, which for this case is given by
\begin{eqnarray}
  \hat{H}_{g}= \sum_{n=0}^{2}\sum_{j=1}^{3}\hat{H}_{g}^{(n,j)},\label{Eq:38}
\end{eqnarray}
where
\begin{eqnarray}
  \hat{H}_{g}^{(n,j)}&=& -\hslash g_{nj} \hat{a}^{(n)\dag}(t)\hat{a}^{(n)}(t) \left(\hat{d}^{(j)}(t)+\hat{d}^{(j)\dag}(t)\right),\label{Eq:OpH2}
\end{eqnarray}
is the optomechanical Hamiltonian for the optical mode $\hat{a}^{(n)}$ coupled to the mechanical vibration mode $\hat{d}^{(j)}$. We may consider three possible modes of oscillation: one longitudinal, with propagation parallel to the wave vector of the field, and two transversal modes. The optomechanical coupling strength $g_{nj}$ is expressed as a frequency. It quantifies the interaction between a single phonon and a single photon.
 The Hamiltonian in Eq. (\ref{Eq:OpH2}) reveals that the interaction of a vibrating non-linear crystal with the radiation field is fundamentally a nonlinear process, involving three operators (three-wave mixing), coupling photon number operators to the creation and annihilation of phonons.

Following a procedure similar to that done in Sec. \ref{hamilt},
we can write the bosonic operator $d^{(j)}$ with the help of the Fourier transform as
\begin{eqnarray}
  \hat{d}^{(j)}(t)= \int_{0}^{\infty}d\Omega_{m} e^{-i\Omega_{m} t} \hat{d}^{(j)}_{\Omega_{m}},\label{Eq:pH1}
\end{eqnarray}
with $\hat{d}^{(j)}_{\Omega_{m}}$ the phonon annihilation operator in the mechanical mode of frequency $\Omega_{m}$. The Hamiltonian in Eq. (\ref{Eq:38}) can also be described by a sum of contributing terms over many different frequencies as
 $ \hat{H}_{g}= \int_{\epsilon}^{\infty}d\Omega \hat{H}_{g}(\Omega)$,
where
\begin{eqnarray}
 \hat{H}_{g}(\Omega)=\sum_{n=0}^{2}\sum_{j=1}^{3}-\hslash g_{nj}\left[ \alpha_{\omega_{n}}\left( \hat{a}_{\omega_{n}-\Omega}^{(n)\dag}\hat{d}_{\Omega}^{(j)\dag}+ \right. \right.\nonumber\\ \left.\left.
\hat{a}_{\omega_{n}+\Omega}^{(n)\dag}\hat{d}_{\Omega}^{(j)} \right)+ \text{h.c.}\right]. \label{Eq:pH3}
\end{eqnarray}
Note that, satisfying energy conservation, different process may occur from the annihilation of a photon of the carrier, described in the linearization by the field amplitude $\alpha_{\omega_{n}}$.
Either we may have the production of a photon in the lower sideband and the production of a phonon from the annihilation of a carrier photon, or the production of a photon in the upper sideband with the annihilation of a phonon. The reverse process are described by the Hermitian conjugate terms.

The complete Hamiltonian of the system, which includes the parametric down conversion and the photon-phonon interaction, would be given by
\begin{eqnarray}
  \hat{H}(\Omega) &=& \hat{H}_{\chi}(\Omega)+\hat{H}_{g}(\Omega),
\end{eqnarray}
 where $\hat{H}_{\chi}(\Omega)$ and $\hat{H}_{g}(\Omega)$ are given by the Eqs. (\ref{a6}) and (\ref{Eq:pH3}), respectively. Now a complete evaluation of the contribution of both parametric down conversion and Brillouin scattering to the OPO dynamics can be performed.

\subsection{Equations of motion for the field quadrature operators}

The evolution of the system should now include the modes of the phonon bath.
Let be $\vec{\mathds{A}}=\left(\vec{\mathbf{A}},\vec{\mathbf{D}}\right)^{T}$, where the field operator vector $\vec{\mathbf{A}}$ was defined in the Sec. \ref{motion} and  $\vec{\mathbf{D}}=(\hat{d}^{(1)}_{\Omega} ~\hat{d}^{(1)\dagger}_{\Omega} ~\hat{d}^{(2)}_{\Omega} ~\hat{d}^{(2)\dagger}_{\Omega} ~\hat{d}^{(3)}_{\Omega} ~\hat{d}^{(3)\dagger}_{\Omega})^{T}$ lists the bosonic operators on the phononic reservoirs.
Therefore the set of differential equations describing the dynamics of operators can be written in compact form as follows:
\begin{eqnarray}\label{Eq:44}
\dfrac{d \vec{\mathds{A}}}{d \xi}=\mathds{M}_{(\chi,g)} \vec{\mathds{A}},
\end{eqnarray}
where
\begin{eqnarray}
\mathds{M}_{(\chi,g)}=\left(\begin{array}{cc}
\mathbf{M}_{\chi} & i\mathbf{J}_{g}\\
i\mathbf{K}_{g} & \mathbf{0}_{6\times6}
\end{array}\right).
\end{eqnarray}
 Here $\mathbf{M}_{\chi}=\mathbf{\Lambda}(\mathbf{M}_{\chi s}\oplus\mathbf{M}_{\chi a})\mathbf{\Lambda}^{-1} $ and
\begin{eqnarray}
  \mathbf{J}_{g} = \left(\begin{array}{c} \mathbf{L} \\
   \mathbf{L}^{\text{'}}
  \end{array}\right),\quad \quad
  \mathbf{K}_{g} = \left(\begin{array}{cc} \mathbf{L}^{\dag} & -\mathbf{L}^{\text{'}\dag}  \end{array}\right),\label{Eq:47}
\end{eqnarray}
where
\begin{eqnarray}
  \mathbf{L}_{nj} &=& g_{nj}\left(\begin{array}{cccccc}
                       \alpha_{\omega_{n}} & 0  \\
                       0 & -\alpha_{\omega_{n}}^{*}
                     \end{array}\right),  \\[2mm]
 \mathbf{L}_{nj}^{\text{'}} &=& g_{nj}\left(\begin{array}{cccccc}
                       0 & \alpha_{\omega_{n}}  \\
                       -\alpha_{\omega_{n}}^{*} & 0
                     \end{array}\right),
\end{eqnarray}
are the elements matrix of the matrices $\mathbf{L}$ and $\mathbf{L}^{\text{'}}$, respectively. In Eq. (\ref{Eq:47}) the "dagger" denotes conjugate transpose of the matrix.

The solution of Eq. (\ref{Eq:44}) is given by
\begin{eqnarray}\label{Eq:48}
\vec{\mathds{A}}\Big| _{\xi=1}=\mathds{G}(\chi,g) \vec{\mathds{A}}\Big|_{\xi=0},
\end{eqnarray}
where
\begin{eqnarray}\label{Eq:49}
\mathds{G}(\chi,g)=\exp\left(\int_{0}^{1}d\xi~\mathds{M}_{(\chi,g)} \right).
\end{eqnarray}

\subsection{Modeling the optical cavity}

Following a procedure similar to that done in Sec. \ref{cavity}, we get similar expressions for the output fields of the cavity. Specifically,
\begin{eqnarray}
\vec{\mathds{A}}_{\textrm{R}}=  \mathds{R}_{(\chi,g)}\vec{\mathds{A}}_{\textrm{in}}+\mathds{T}'_{(\chi,g)}
\vec{\mathds{A}}_{\nu}.\label{Eq:50}
\end{eqnarray}
The expressions for the matrices $\mathds{R}_{(\chi,g)}$ and $\mathds{T}'_{(\chi,g)}$ are similar to those given in Eqs. (\ref{Eq:26}) and (\ref{Eq:27}) but with the following modifications to account for
 the phonon operators.
\begin{align}
  \boldsymbol{\varphi} &\rightarrow \Psi = \left( \boldsymbol{\varphi}\oplus \mathbf{0}_{6\times6} \right), \notag\\
  \mathbf{R} &\rightarrow \mathds{R} = (\mathbf{R}\oplus\mathbf{0}_{6\times6}), \notag\\
   \mathbf{T} &\rightarrow \mathds{T} = (\mathbf{T}\oplus\mathbf{1}_{6\times6}), \notag\\
  \mathbf{R}' &\rightarrow \mathds{R}' = (\mathbf{R}'\oplus\mathbf{0}_{6\times6}), \notag\\
  \mathbf{T}' &\rightarrow \mathds{T}' = (\mathbf{T}'\oplus\mathbf{1}_{6\times6}).\notag
\end{align}

\subsection{Solution for the Gaussian quantum state: covariance matrix in the eigenbasis of quadrature operators}

In analogy with the Eq. (\ref{Eq:30}), the covariance matrix for all fields (optical and phononic) is
\begin{equation}
\mathds{V}_{\textrm{R}}=\widetilde{\mathds{R}}_{(\chi,g)}\mathds{V}_{\text{in}}
\widetilde{\mathds{R}}_{(\chi,g)}^{T}+ \widetilde{\mathds{T}}_{(\chi,g)}^{'}\mathds{V}_{\nu} \widetilde{\mathds{T}}_{(\chi,g)}^{'T}.
\label{Eq:50b}
\end{equation}
Considering the case where field inputs are in vacuum state, and the phonon reservoir is in a thermal state, $\mathbf{V}_{\text{th}}=(1+2\bar{n}_{\text{th}})\mathbf{1}_{6\times6}$, we have
\begin{equation}\label{Eq:51}
  \mathds{V}_{in}=\mathds{V}_{\nu}=\left( \mathbf{1}_{12\times12}\oplus\mathbf{V}_{\text{th}}\right),
\end{equation}
considering here that the three phonon modes of the reservoir have the same temperature and the same average number of phonons $\bar{n}_{\text{th}}$.

The resulting covariance matrix will be given by
\begin{align}
\mathbf{V}_{s}=
\left(
  \begin{array}{cccccc}
    \rho^{(0)} & e_{1} & \mu^{(01)} & e_{2} & \mu^{(02)} & e_{3} \\
    e_{1} & \beta^{(0)} & e_{4} & \nu^{(01)} & e_{5} & \nu^{(02)} \\
    \mu^{(01)} & e_{4} & \rho^{(1)} & e_{6} & \zeta^{(12)} & e_{7} \\
    e_{2} & \nu^{(01)} & e_{6} & \beta^{(1)} & e_{8} & \epsilon^{(12)} \\
    \mu^{(02)} & e_{5} & \zeta^{(12)} & e_{8} & \rho^{(2)} & e_{9} \\
    e_{3} & \nu^{(02)} & e_{7} & \epsilon^{(12)} & e_{9} & \beta^{(2)}
  \end{array}
  \right),\label{covarphT1}
\end{align}
and
\begin{align}
\mathbf{C}_{s/a}=
\left(
  \begin{array}{cccccc}
    \delta^{(0)}  & 0 &  h_{1} & -\kappa^{(01)} &  h_{2} & -\kappa^{(02)} \\
    0 & \delta^{(0)}  & \lambda^{(01)} &  h_{3} & \lambda^{(02)} &  h_{4} \\
     h_{3} & \kappa^{(01)} & \delta^{(1)}  & 0 &  h_{5} & -\varrho^{(12)} \\
    -\lambda^{(01)} &  h_{1} & 0 & \delta^{(1)}  & \eta^{(12)} &  h_{6} \\
     h_{4} & \kappa^{(02)} &  h_{6} & \varrho^{(12)} & \delta^{(2)}  & 0 \\
    -\lambda^{(02)} &  h_{2} & -\eta^{(12)} &  h_{5} & 0 & \delta^{(2)}
  \end{array}
  \right).\label{covarphT2}
\end{align}

A direct comparison with matrices in Eqs. (\ref{covarsphT1},\ref{covarsphT2}) shows many additional features coming from this added thermal reservoir. It is curious that even in the absence of phonons in the reservoir, those terms should yet appear due to the photon-phonon coupling of the zero-temperature fluctuations.
Nevertheless, these terms will be small in this case, and would not affect significantly the covariance, even though the resulting state of the field is no longer pure due to the coupling to extra modes from the crystal.

\section{Experimental results \label{experiment}}

\begin{figure}[h]
  \begin{center}
    \includegraphics[width=0.8\columnwidth]{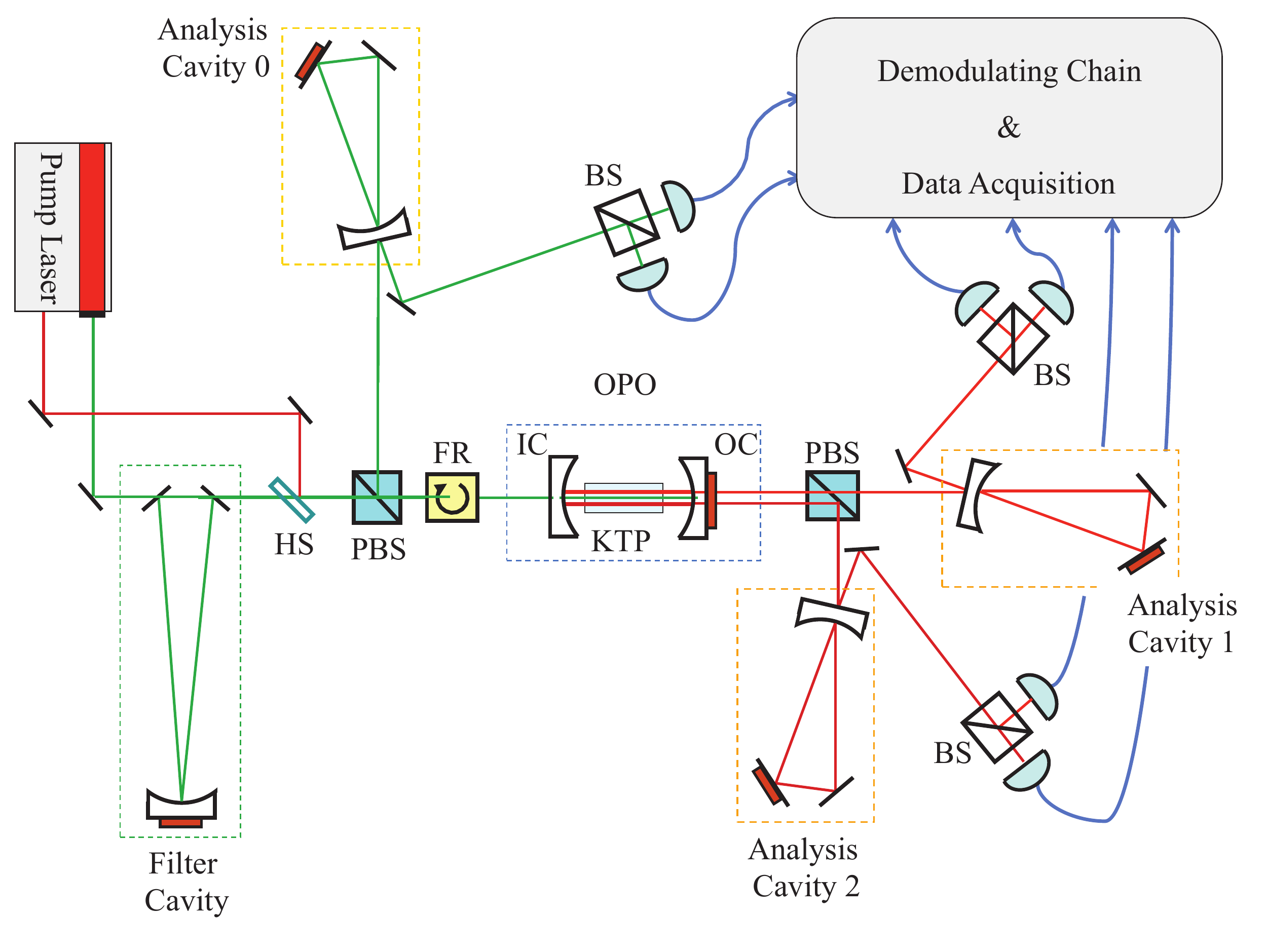}
    \caption{(Color online) Setup for the reconstruction of the OPO beams' covariance matrix. PBS, polarizing beam splitter; BS, 50:50 beam splitter; HS, harmonic separator; IC, input coupler; OC, output coupler (OPO cavity); FR, Faraday rotator.}
    \label{fig:setup}
  \end{center}
\end{figure}

The model developed here can be directly compared to the experimental results obtained from the setup described in \cite{hexaopo}.
The system is a triply resonant OPO operating above threshold, and the experimental setup is depicted in Fig. \ref{fig:setup}. The OPO cavity is pumped by the second harmonic of a doubled Nd:YAG laser, filtered with a mode cleaning cavity to ensure that pump fluctuations are reduced to the standard quantum level in amplitude and phase for frequencies above 20 MHz.

\begin{figure}[h]
\centerline{\includegraphics[width=\columnwidth]{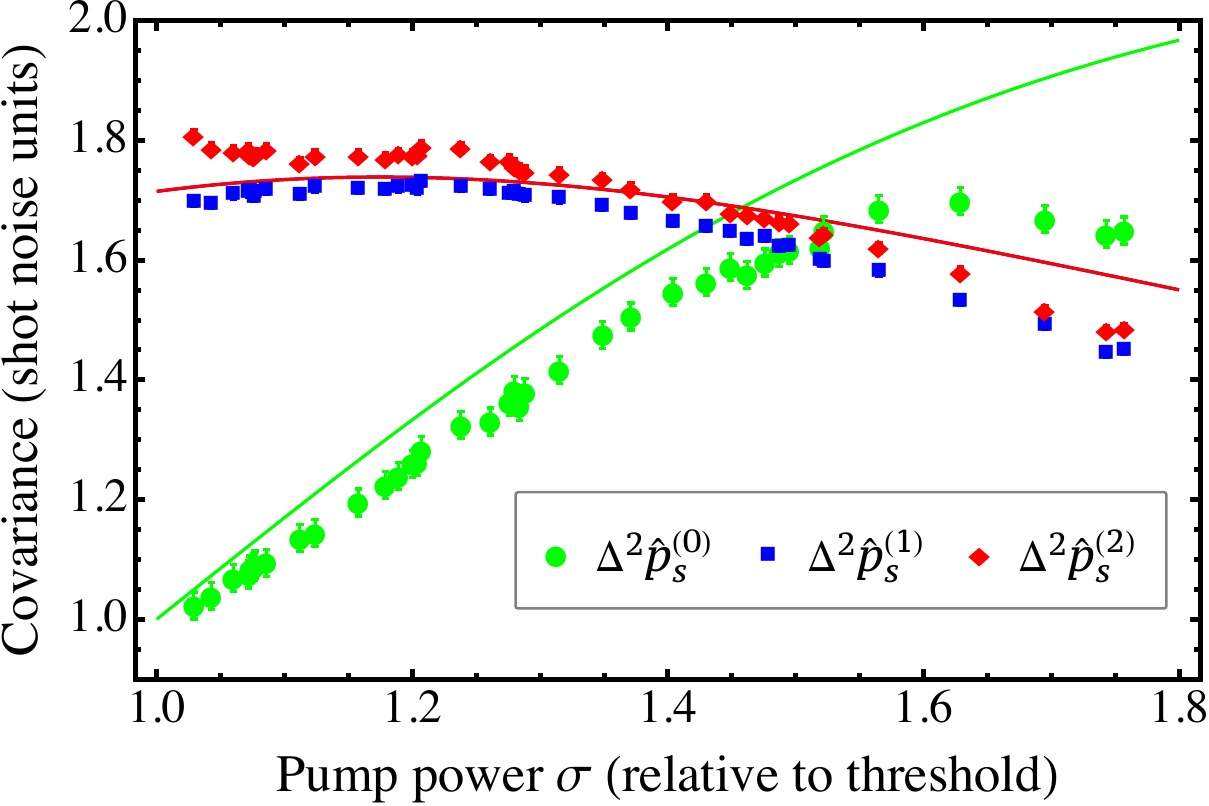}}
\centerline{\includegraphics[width=\columnwidth]{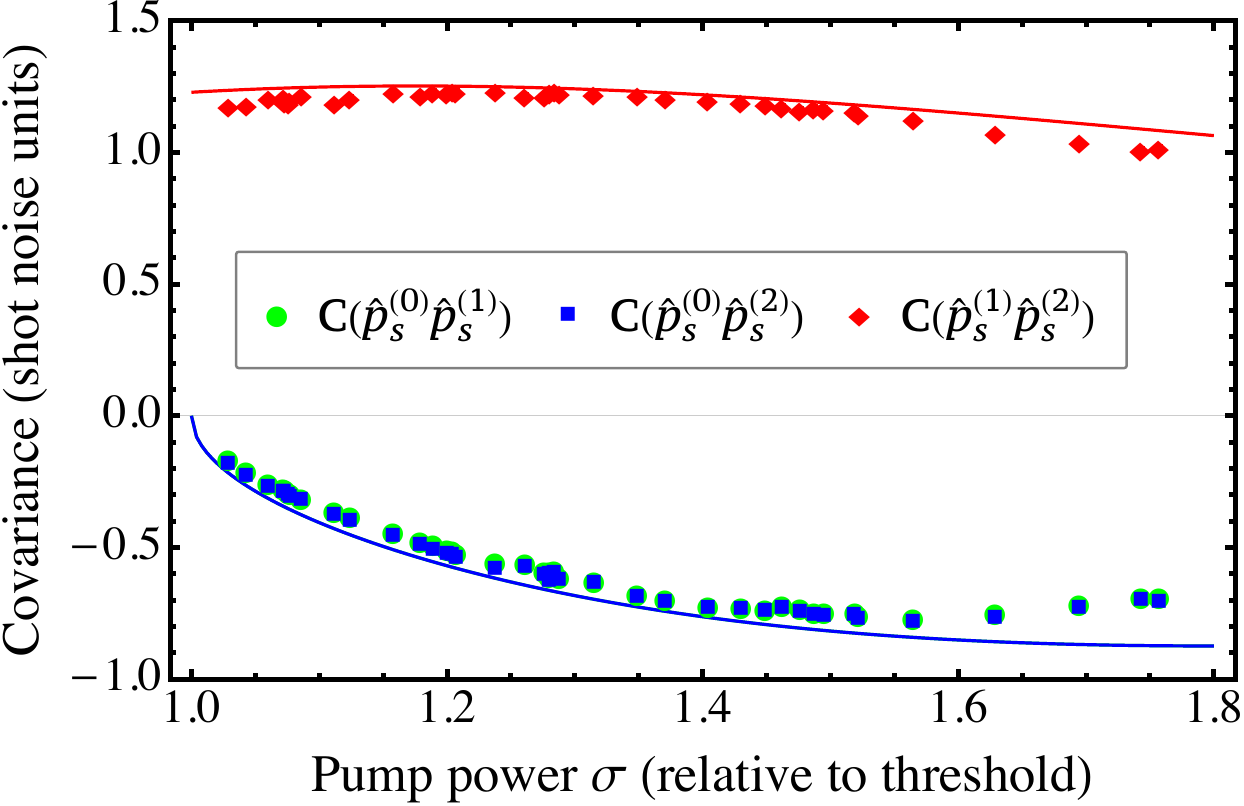}}
\centerline{\includegraphics[width=\columnwidth]{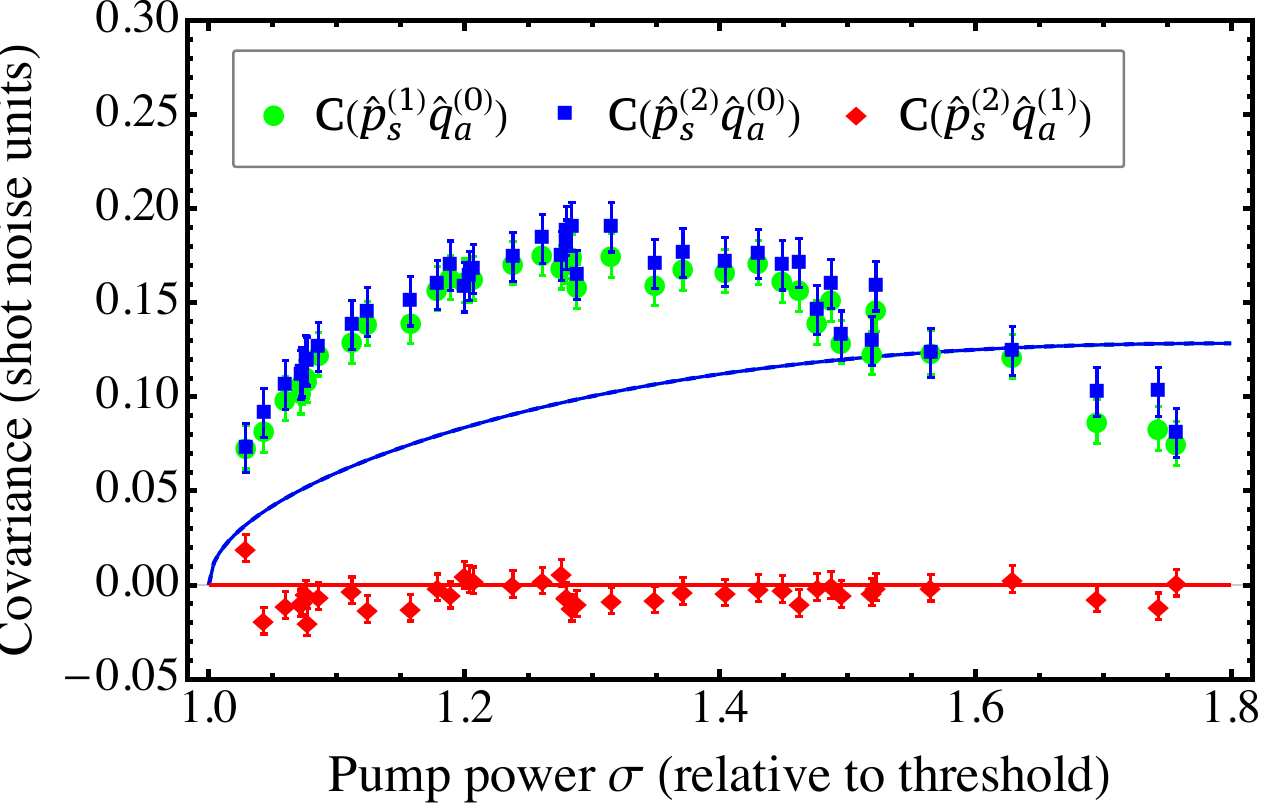}}
\caption{Measured variances of the amplitudes of the three fields coming from the OPO, in the symmetric description, followed by their respective correlations. Cross correlations between symmetric and antisymmetric modes. } \label{fig:covarP}
\end{figure}

\begin{figure}[h]
%\centering
\centerline{\includegraphics[width=\columnwidth]{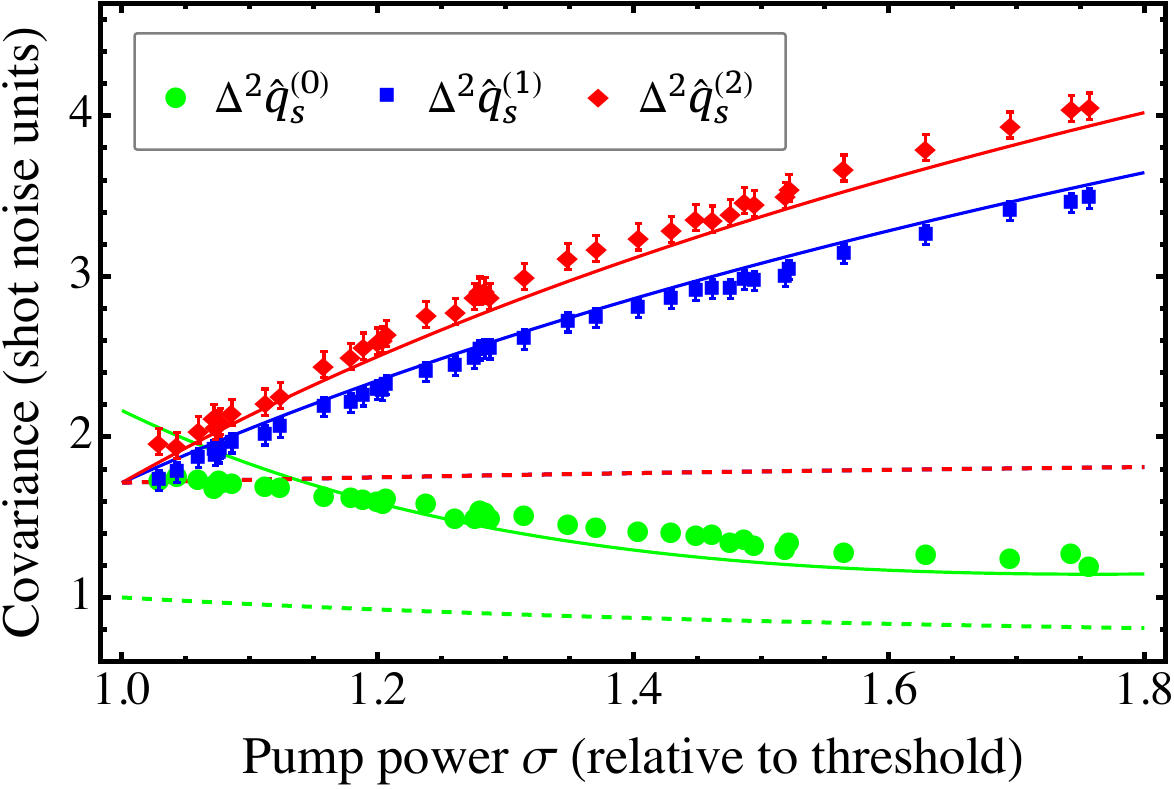}}
\centerline{\includegraphics[width=\columnwidth]{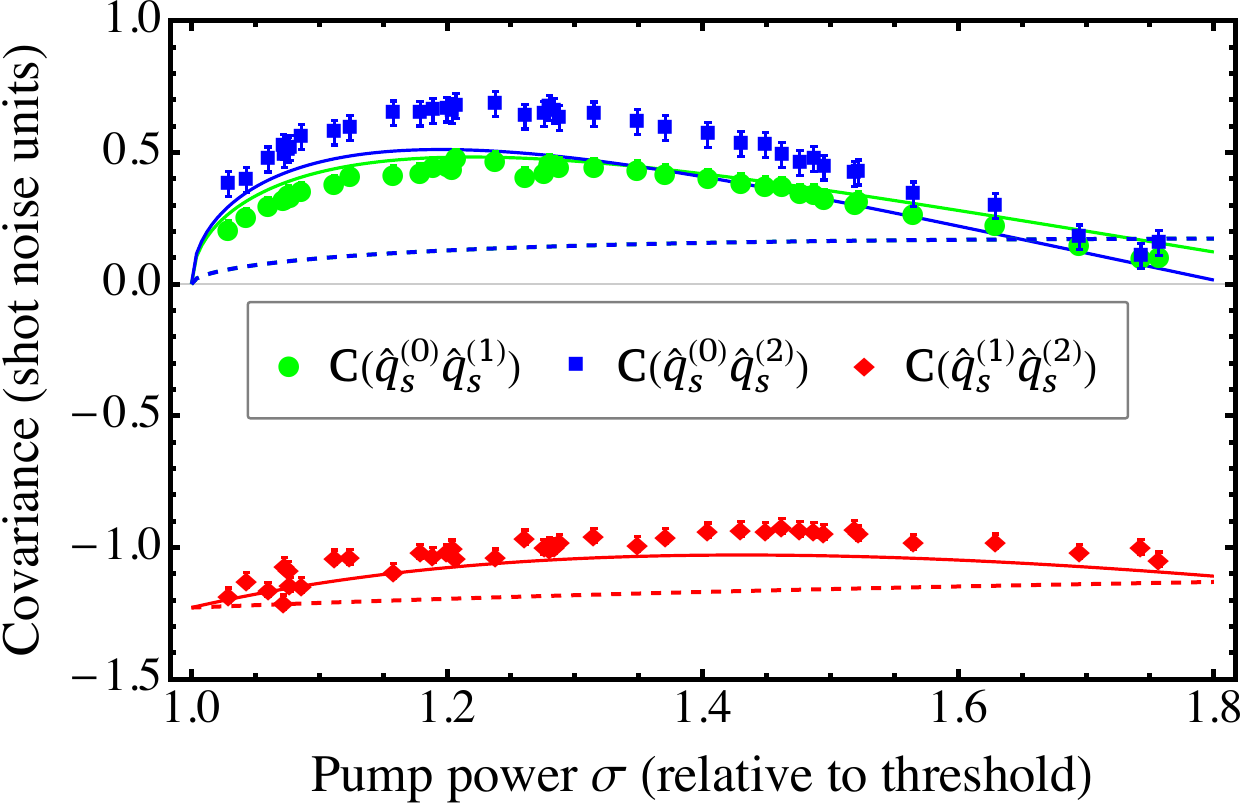}}
\centerline{\includegraphics[width=\columnwidth]{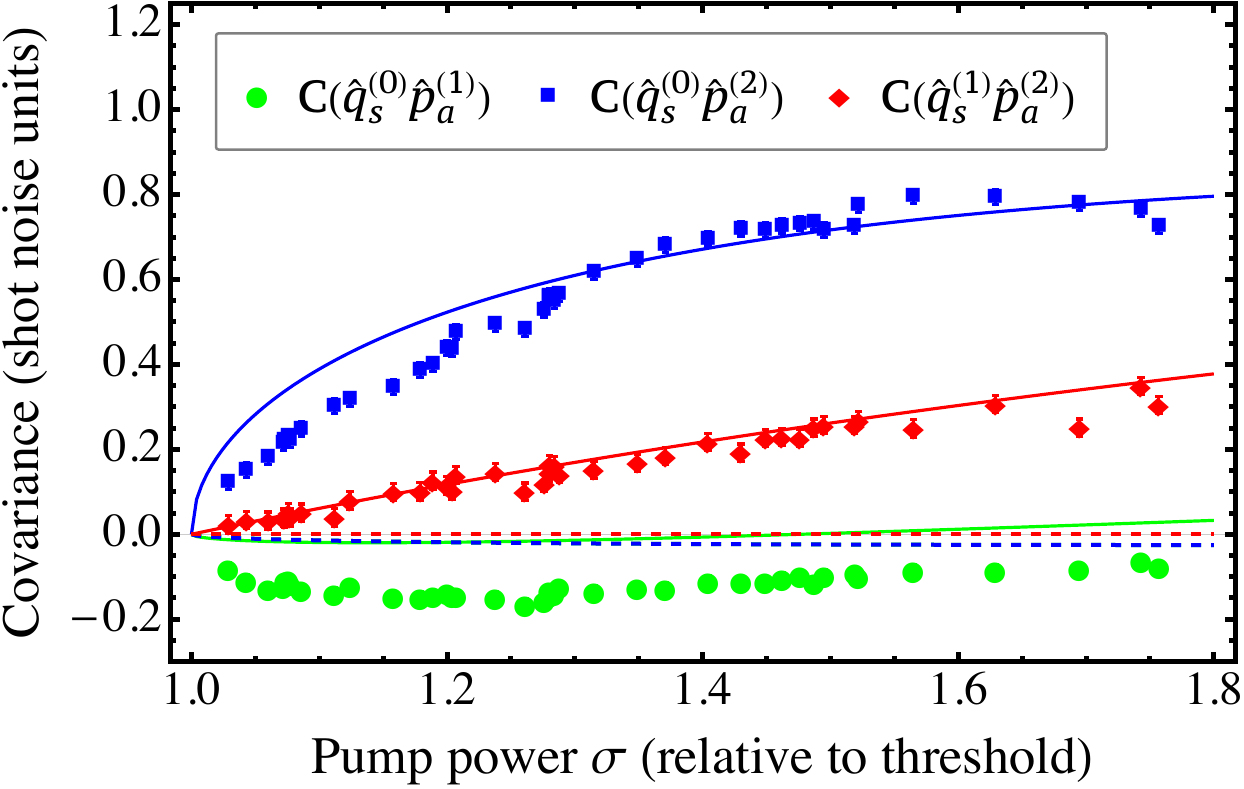}}
\caption{Measured variances of the phase of the three fields coming from the OPO, in the symmetric description, followed by their respective correlations. Cross correlations between symmetric and antisymmetric modes. Dashed lines are the result we would expect in the absence of phonons noise. } \label{fig:covarQ}
\end{figure}

\begin{figure*}[t]
\centering
\includegraphics[width=0.45\linewidth]{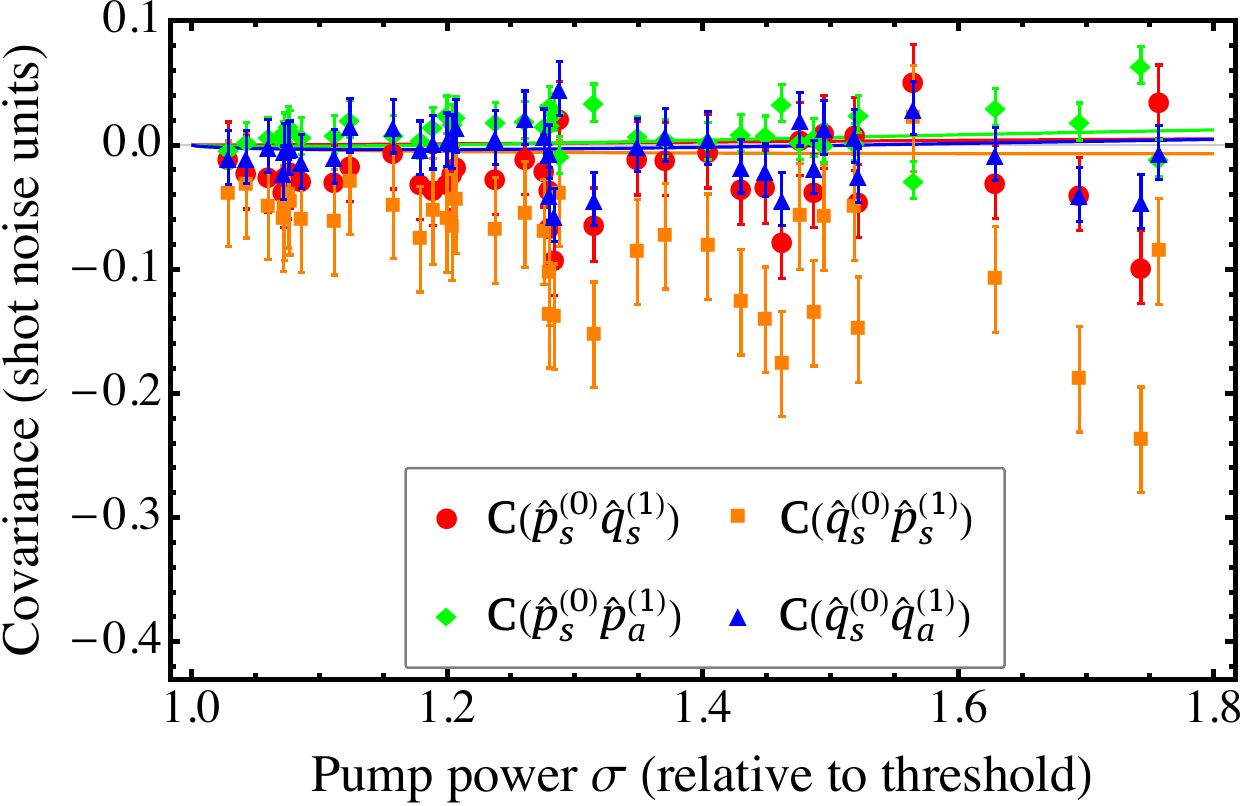}
\includegraphics[width=0.45\linewidth]{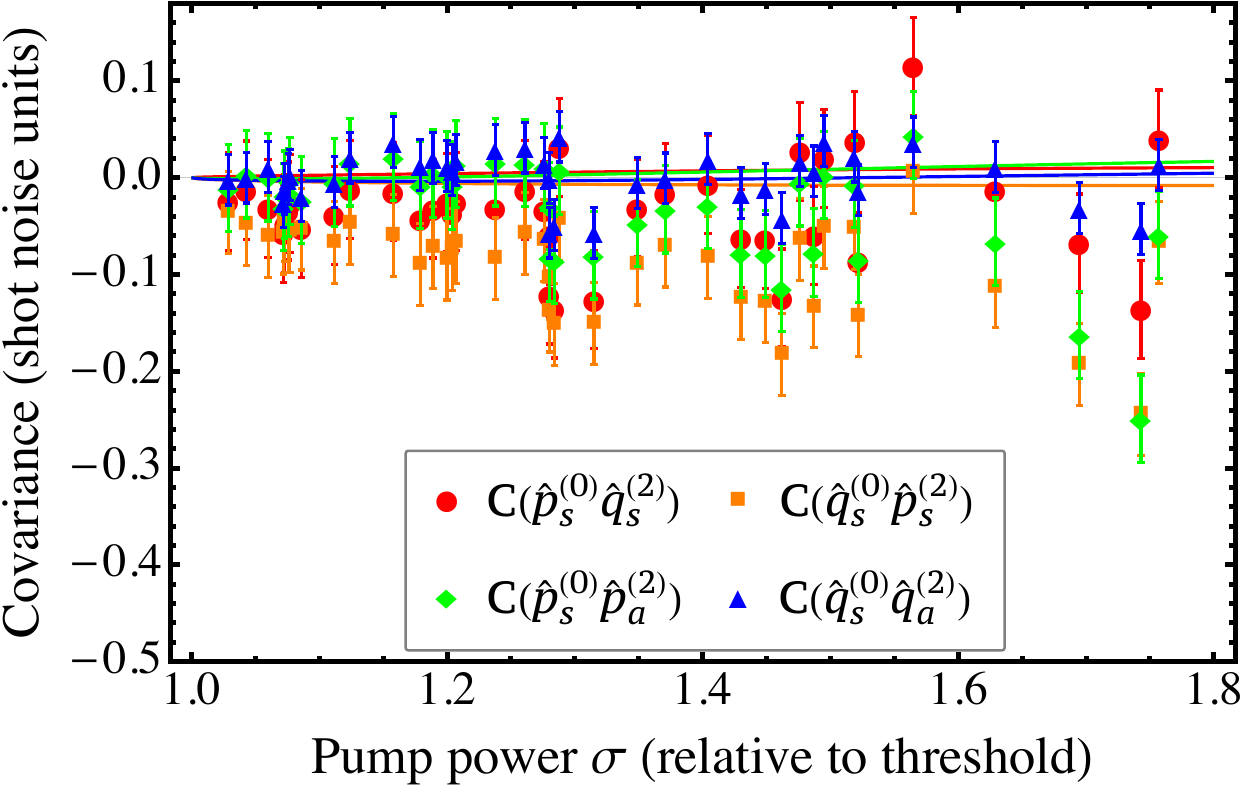}
\includegraphics[width=0.45\linewidth]{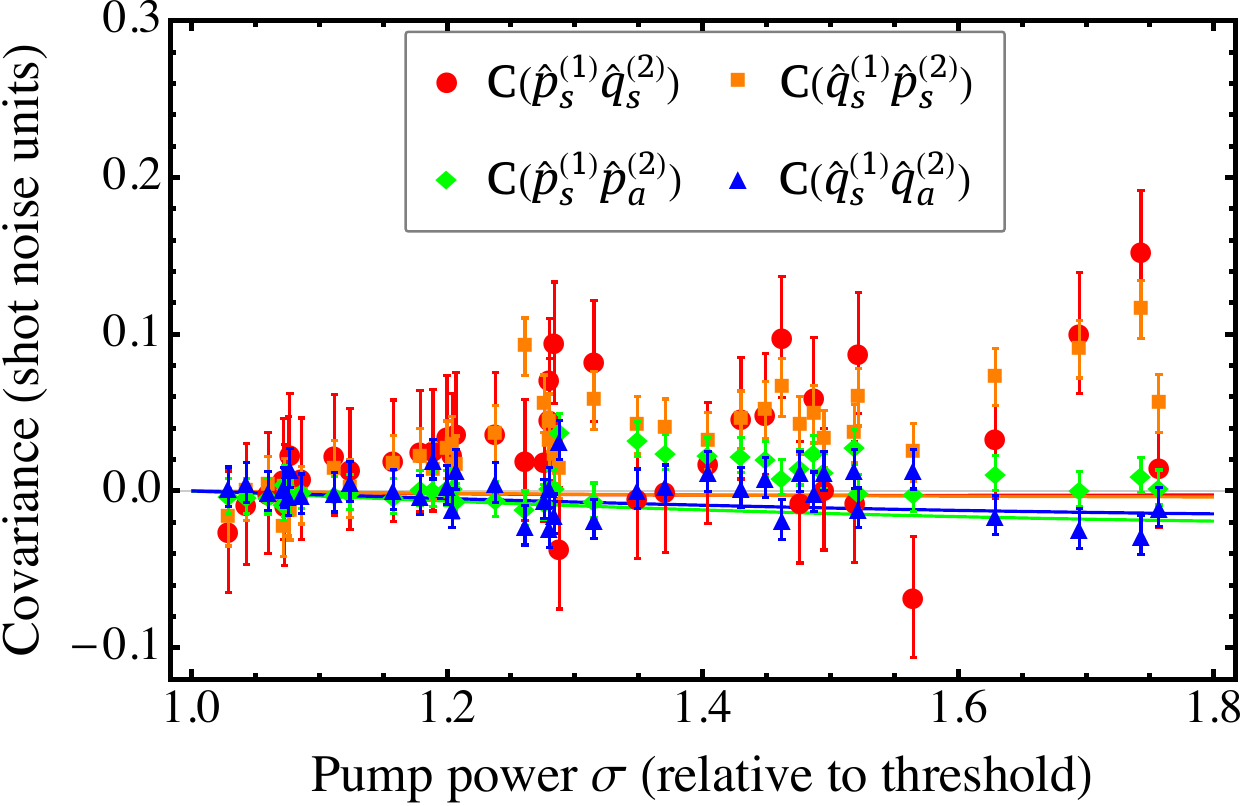}
\includegraphics[width=0.45\linewidth]{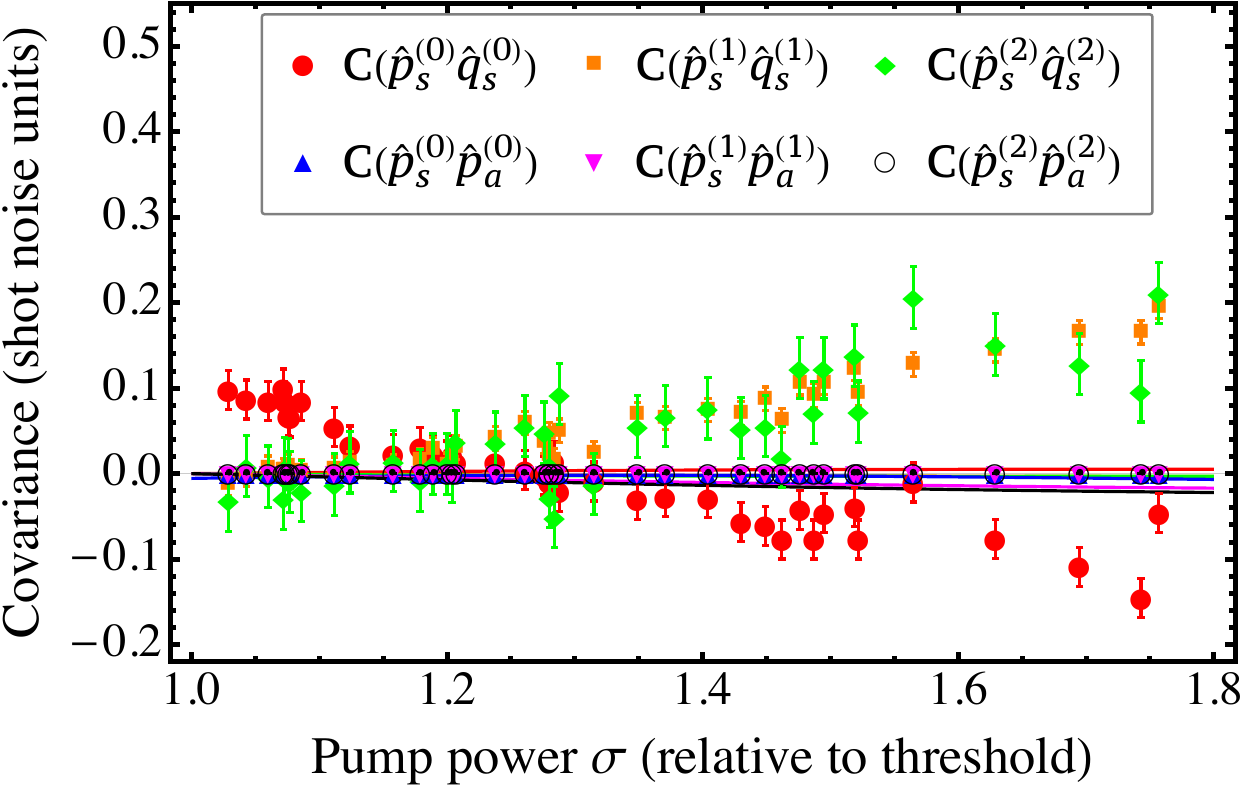}
\caption{Measured correlations between amplitude and phase for each mode in symmetric/antisymmetric description. } \label{fig:covarPQ}
\end{figure*}

The filtered pump beam is then injected in the OPO, with adjustable power, through the input coupler (IC) with a reflectivity of 70$\%$ for the pump field (532 nm) and high reflectivity ($>99\%$) at 1064 nm. The reflected pumped field is recovered from the Faraday rotator (FR). The infrared output coupler (OC) has a reflectivity of 96$\%$ at $\approx$1064 nm and high reflectivity ($>99\%$) at 532 nm. Both mirrors are deposited on concave substrates with a curvature radius of 50 mm. The crystal is a type II phase-matched KTP (potassium titanyl phosphate, KTiOPO$_{4}$) with length $l=12$ mm, average refractive index n=1.81(1) and antireflective coatings for both wavelengths. The average free spectral range for the three modes is found to be of 4.3(5) GHz. The cavity finesse for pump mode is 15 and 124 for the signal and idler modes (the latter defined as the mode with the same polarization as the pump). The overall detection efficiencies are 87$\%$ for the infrared beams and 65$\%$ for the pump, accounting for detector efficiencies and losses in the beam paths. The threshold power is 60 mW, and the maximum pump power was 75\% above the threshold. In order to reduce the effect of phonon noise on the system, the crystal is cooled to 260 K, and the OPO is kept in a  vacuum chamber to avoid condensation.

Phase noise measurements were performed using the ellipse rotation method described in \cite{Galatola1991,Villar2008}, with the help of analysis cavities. Cavities 1 and 2 (for the transmitted infrared beams) have bandwidths of 14(1) MHz, and cavity 0 (for the reflected pump) has a bandwidth of 12(1) MHz. This ensures a full rotation of the noise ellipse 
for the chosen analysis frequency of 21 MHz. Mode matching of the beams to the analysis cavities was better than 95$\%$. Combining in-quadrature electronic local oscillators and cavity detection \cite{hexaopo, prlsideband}, we were able to reconstruct the covariance matrix of the output sidebands. Since the detected modes are of Gaussian nature \cite{gaussian}, determination of the covariance matrix is equivalent to the complete tomography of the output state of the sidebands of the intense optical fields involved.

Covariances for the intensity fluctuations are shown in Fig. \ref{fig:covarP}, in terms of the symmetric/antisymmetric modes, that results in a compact presentation of the covariance matrix. They present a good agreement of the theory and the experiment. Deviations for the pump field at higher pump power are consistent with the effects of mismatch in the pumping field, that are aggravated by thermal lensing of the crystal. The pump cavity mode will be more depleted with growing pump power, and the contributions of unmatched modes will be more relevant, degrading the measurement of the variance and contributing as an effective loss in detection. Nevertheless, correlations are less affected in this case, and present a better agreement. It is curious to notice that  correlations between the symmetric and antisymmetric modes are observed for pump and signal (or idler) correlations, as predicted in \cite{hexaopo}, revealing that there is more information on the system beyond the three mode description. A full description of the measured state should necessarily involve six fields, and the distinct role of each sideband becomes relevant for the tomography of the system.

Phase quadrature measurements of fields of distinct colors are possible by the use of analysis cavities. The results shown in Fig. \ref{fig:covarQ} were evaluated with a limited number of adjusting variables to describe the phonon coupling. The complete model involves three coupling channels between each mode of the carrier to distinct reservoirs, one for each oscillating mode of a crystal. Nevertheless, a toy model considering that pump and idler are coupled to the same reservoir (since they have the same polarization), and the orthogonally polarized signal with additional coupling to a second reservoir can be used to adjust the curves to the data. Best results were obtained with $g_{01}= 8.0 \, 10^{-3}$ for the pump coupling, $g_{21}= 3.6 \, 10^{-3}$ for the idler coupling, and $g_{11}=1.9 \,10^{-3}$ for signal coupling to one of the reservoirs, and $g_{12}=2.7 \, 10 ^{-3}$ for signal coupling to the second reservoir. Thermal phonon population density was arbitrarily set to $N_{th}=100$, acting just as a multiplicative constant in our model at high temperatures. It is curious to notice that $\sqrt{g^2_{11}+g^2_{12}}\sim g_{21}$, and $g_{01}\sim 2 g_{21}$, consistent with the scaling with wavelength described in the semiclassical model for the phonon noise \cite{phononpra}.

It is clear that the photon-phonon coupling leads to an additional noise to the system, that should degrade the purity, if we compare with the expected results of the variance in absence of phonon noise, shown by dashed lines in Fig. \ref{fig:covarQ}. This coupling prevents the observation of phase squeezing for the pump mode in the present condition, and adds noise to signal and idler fields. Since this additional noise is not perfectly correlated, it will lead to degradation on the squeezing level at the sum of the phases, as we would expect in the generation of entangled modes of the field \cite{reiddrummond}. Nevertheless, quantum correlations for two \cite{villarentangled}  and three modes \cite{tripartite} can be observed if adequate control of the phonons is available, by the cooling of the crystal.

So far, we have presented all the measurements for the 18 terms on matrices given in Eqs. (\ref{covarphT1}) and (\ref{covarphT2}). Nevertheless, a complete description of the system should involve all the correlations between phase and amplitudes of each field in symmetric and antisymmetric description. The present model shows that for perfect resonance of the carriers the contribution of these terms should be zero. Experimental results are close to this situation for low pump powers, as can be seen in Fig. \ref{fig:covarPQ}. Cross-correlations become effectively nonzero for growing pump powers, where thermal effects should provide some change in the refractive index leading to small detunings of the carrier modes.

\section{Conclusion\label{conclusion}}

In continuous variables domain, the combined use of self-homodyning \cite{Villar2004} and demodulation by in-quadrature local oscillators \cite{hexaopo} allows the complete reconstruction of the state of six modes of the field in an above threshold OPO. These modes are related to the sidebands of the downconverted fields, generated by the nonlinear process, and the pump field, reflected by the cavity.
The results we obtained are in good agreement with the detailed model developed here, involving the transformation of the field operators in their reflection by a cavity,  the nonlinear coupling among the fields by the crystal and the photon-phonon coupling.
For the linear approach we had chosen, the model reproduces the so-called ``semiclassical model" of the OPO, where quantized fields can be associated to stochastic fluctuations in a Langevin equation, leading to a spectral matrix, associated with the Fourier transform of the two-time correlation of the output fields. In the present case, discrepancies between our model and the semiclassical one are smaller than 4\% of the standard quantum level (except for amplitude variance of the pump, reaching 9\%) being both compatible with the experimental results. 

The main result of the developed model is the demonstration that the imaginary part of the spectral matrix, i.e. the correlations between symmetric and asymmetric combinations of sidebands \cite{hexaopo}, has not its physical origin in the nonlinear process but on the evolution of the fields inside the cavity, combined with the effective beam splitter transformation for downconverted and pump modes, explicit derived in the linearized model. This particular effect is not explicit in the semiclassical treatment.
The asymmetries in phase evolution of upper and lower sidebands lead to the coupling of their symmetric and asymmetric combinations. These effects will be small for reduced analysis frequencies and will be maximized as they get closer to the OPO cavity bandwidth.

The presented model is shown to be suitable for the reconstruction of the covariance matrix in a linearized regime, valid for small intracavity gain. Since the output fields are in a Gaussian state for all practical purposes \cite{gaussian},  it characterizes a complete state tomography involving six modes of an OPO in a wide range of pump values, opening the path to explore the structure of hexapartite entanglement in this system \cite{hexapartite}.

%{\color{red}The model should be valid even in an open cavity regime,  and can be applied to situations where the cavity is completely open for some of the fields involved \cite{opencavity}. Since it deals with the transformation of the fields, it can be used beyond the linearization of the intracavity fields, eventually showing non-Gaussian behaviors in nonlinear processes.}
%As we had demonstrated, all elements of this matrix can be measured \cite{hexaopo}, and the imaginary part of this Hermitian matrix is associated with the correlation between symmetric and asymmetric combinations of the sidebands.
%The model we have developed here demonstrate that the origin of this imaginary term is not in the nonlinear process, but on the evolution of the fields inside the cavity. Asymmetries in phase evolution of upper and lower sidebands lead to the coupling of their symmetric and asymmetric combinations. These effects will be small for reduced analysis frequencies, and will be maximized as the get closer to the OPO cavity bandwidth.

\section{Acknowledgments}

The authors acknowledge the support from grant \# 2010/08448-2, \href{http://dx.doi.org/10.13039/501100001807}{Funda\c c\~ao de Amparo \`a Pesquisa do Estado de S\~ao Paulo (FAPESP)},
\href{http://dx.doi.org/10.13039/501100003593}{Conselho Nacional de Desenvolvimento Cient\'\i fico e Tecnol\'ogico}, and
Coordena\c c\~ao de Aperfei\c coamento de Pessoal de N\'\i vel Superior

%\bibliography{Draft_PRA_OPO}% Produces the bibliography via BibTeX.

\end{document}